# Reconfigurable nonlinear optical computing device for retina-inspired computing

Xiayang Hua[1], Jiyuan Zheng*[1,2]✉, Peiyuan Zhao[3], Hualong Ren[4], Xiangwei Zeng[4], Zhibiao Hao[1,2], Changzheng Sun[1,2], Bing Xiong[1], Yanjun Han[1], Jian Wang[1], Hongtao Li[1], Lin Gan[1], Yi Luo[1,2] and Lai Wang*[1,2]✉

[1]Department of Electronic Engineering, Tsinghua University, Beijing 100084, China.
[2]Beijing National Research Center for Information Science and Technology (BNRist).
[3]School of Software, Tsinghua University.
[4]Neurocean Technologies Inc.
[5]These authors contributed equally: Xiayang Hua, Jiyuan Zheng, Peiyuan Zhao.
✉ e-mail: zhengjiyuan@tsinghua.edu.cn; wanglai@mail.tsinghua.edu.cn;

**Abstract**

Optical neural networks are at the forefront of computational innovation, utilizing photons as the primary carriers of information and employing optical components for computation. However, the fundamental nonlinear optical device in the neural networks is barely satisfied because of its high energy threshold and poor reconfigurability. This paper proposes and demonstrates an optical sigmoid-type nonlinear computation mode of Vertical-Cavity Surface-Emitting Lasers (VCSELs) biased beneath the threshold. The device is programmable by simply adjusting the injection current. The device exhibits sigmoid-type nonlinear performance at a low input optical power ranging from merely 3-250 μW. The tuning sensitivity of the device to the programming current density can be as large as 15 μW*mm$^2$/mA. Deep neural network architecture based on such device has been proposed and demonstrated by simulation on recognizing hand-writing digital dataset, and a 97.3% accuracy has been achieved. A step further, the nonlinear reconfigurability is found to be highly useful to enhance the adaptability of the networks, which is demonstrated by significantly improving the recognition accuracy by 41.76%, 19.2%, and 25.89% of low-contrast hand-writing digital images under high exposure, low exposure, and high random noise respectively.

## INTRODUCTION

In recent years, the rapid progress in artificial intelligence and the advent of Large Language Models (LLMs) have underscored the urgent need for chips that can process information more swiftly and efficiently, while also reducing power consumption[1]. This demand presents a formidable challenge, as traditional silicon-based chips are approaching their physical limitations. Theoretically, the substitution of electrons by photons for information transmission and computation could offer significant improvements in both speed and energy efficiency. Since the advent of lasers, researchers have been exploring the realm of optical computing. Currently, solutions for optical linear matrix operations are well-established.[2-14] However, achieving nonlinear operations akin to neuronal activation in the optical domain, which is crucial for enhancing the expressive capabilities of optical neural networks (ONNs), remains a significant hurdle. Thus, the development of optical devices that can perform nonlinear activation functions is pivotal for the realization of all-optical neural networks.

In current research, nonlinear activation functions are predominantly implemented in the electrical domain, necessitating the integration of additional components such as converters, optoelectronic and electro-optical sensors, and ancillary driving circuits. These additions inevitably complicate the system, increase latency, and raise power consumption. Consequently, achieving nonlinear activation in the optical domain is essential for the advancement of all-optical neural networks.

Thus far, many research teams have been committed to proposing unique solutions to achieve nonlinear activation in the optical domain. In 2019, B. Shi *et al.* reported on an all-optical nonlinear function leveraging the wavelength-shifting capability of a single nonlinear Semiconductor Optical Amplifier (SOA).[15] In 2021, J. Crnjanski *et al.* introduced an injection locked Fabry Perot (F-P) laser that can perform nonlinear activation of pulsed light similar to PReLU[16]. In 2022, Zhang *et al.* proposed a runway-type Multimode Resonator (MRR) as all-optical nonlinear activator, which exploited the material's thermo-optic properties in the 1550 nm band to realize distinct types of nonlinear functions at an impressively low threshold of 0.75 mW[17]. In 2023, Xiang *et al.* proposed that Fabry-Pérot lasers with saturated cavities (FP-SA) could achieve ReLU-like nonlinear integration for time-domain pulsed light[18].

Despite these advancements, all-optical nonlinear activation schemes still faces several challenges that may hinder their broader application: Firstly, existing optical nonlinear activation schemes typically require high input power, accounting for the high energy consumption[19]. Secondly, most of existing schemes are designed for pulsed light inputs, limited to spike neural network (SNN) algorithms[20,21]. Finally, the limited reconfigurability of nonlinear functions often hampers the adaptability and scalability of optical computing systems, posing a significant limitation in their practical application. [22,23] The importance of this reconfigurability is understandable with a common example. As shown in Fig. 1a, glare may occur when we drive out of the tunnel, which could be dangerous. The reconfigurable visual processing system can quickly adapt to the changes of ambient brightness.[24,25] Fig. 1b shows the traditional image processing architecture with fixed nonlinear characteristics. As shown in Fig. 1c, the retina of the human eye has differentiated sensing cells with different nonlinear characteristics, making our eyes quickly adapt to external variations. Inspired by the retina, we demonstrate a simple and configurable optical nonlinear activation based on VCSEL as shown in Fig. 1d. During the operation, the bias current of VCSEL is set beneath the threshold, which allows the VCSEL to provide reflective gain to incident light near 1550 nm with little excitation. There is a significant nonlinear characteristic between the output power and the incident power.

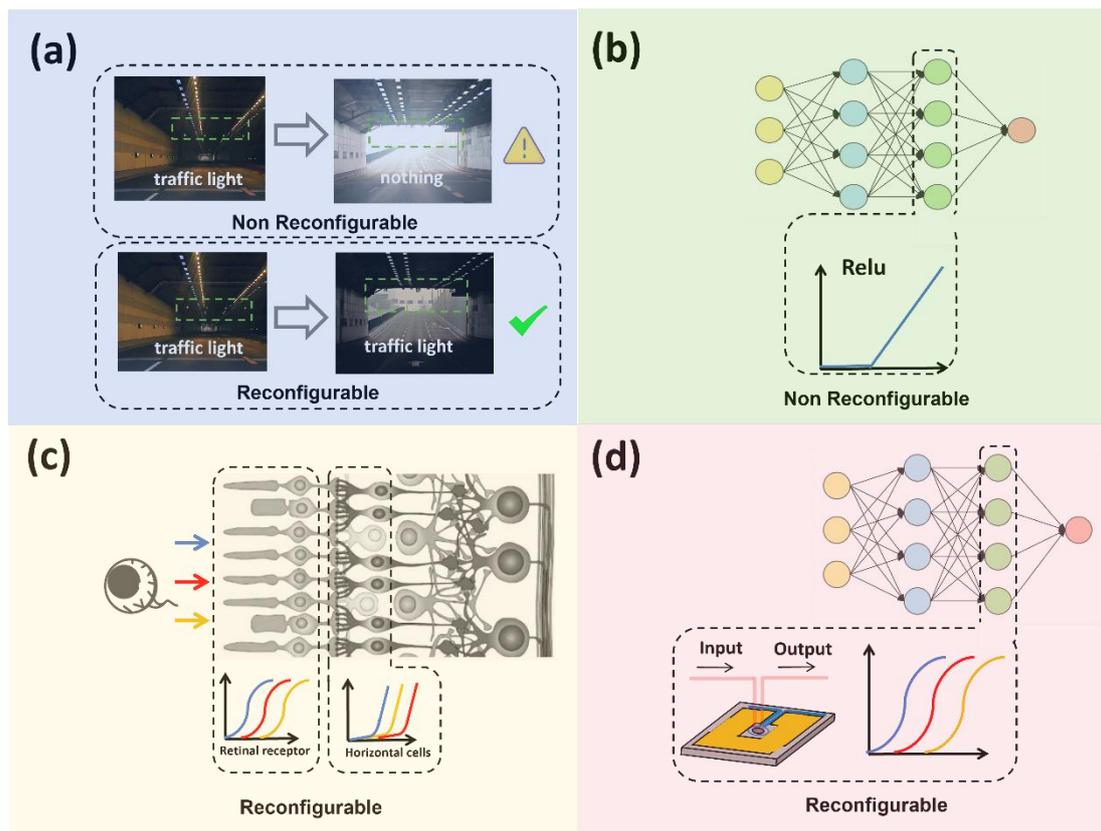

**Fig. 1 The role of reconfigurable nonlinearity in adaptive intelligent computing.** (a) Taking the adaptive recognition ability of traffic lights by imaging devices traversing a tunnel as an example to illustrate the function of reconfigurable nonlinearity: Reconfigurable systems can quickly adapt to bright external environment to avoid glare. (b) The traditional visual neural networks are non-reconfigurable with consistent nonlinear functions[26]. (c) The human retinal system, with its reconfigurable nonlinear characteristics, could process image information adaptively[27]. (d) Inspired by the human retina, this work carried out research on reconfigurable nonlinear computing devices and neural networks based on VCSEL.

**RESULTS**

In the following section, we present our experimental setup and elucidate the underlying physical principles responsible for the nonlinear behavior of our device. As shown in Fig. 2a, the input light is generated from the tunable laser (~1550 nm), after passing through the isolator, it is injected via port 1 into the VCSEL, and the reflected light is exported from port 3. The circulator can effectively separate the incident light and reflected light. The threshold current density of our VCSEL is 175 mA/mm², so the bias current density is set from 150 mA/mm² to 170 mA/mm². With light injected, two peaks can be observed at the output port of the circulator with a spectrometer as shown in Fig. 2b. Peak 1 is the amplified incident light, while Peak 2 comes from the spontaneous emission of VCSEL. Besides, the determination of the gain peak is very important because the gain of incident light depends on its detuning with the gain peak. When we calibrate the gain peak with a low input power (1 μW), its position often coincides with the VCSEL emission peak as shown in Fig. 2c. However, as the input power increases, the position of the gain peak will undergo a significant blue shift, as shown in Fig. 2d, while the VCSEL's emission peak will undergo a slight blue shift, as shown in Fig. 2e. It is known that the emission wavelength of VCSEL is mainly determined by the structure of DBRs, while the gain spectrum of the VCSEL is primarily ruled by the active region. When external light is injected into VCSEL, its physical process can be described by formula (1).[28]

$$\frac{dN}{dt} = \frac{\eta I}{e\Gamma_1 V} - (AN + BN^2 + CN^3) - \frac{\Gamma c \zeta a (N - N_0)}{n_c}(\beta S_a + S_{in}) \tag{1}$$

Where $N$ is the carrier concentration in the active region, $I$ is the bias current, $S_a$ is the intracavity spontaneous emission intensity, and $S_{in}$ is the injection light intensity. When the device operates steadily ( $dN=0$ ), the injected charge carriers are mainly consumed by radiative recombination($AN$), non-radiative recombination($BN^2$), Auger recombination ($CN^3$), and stimulated recombination which is mainly caused by the inject light[29]. These physical processes compete with each other because the injected current remains constant. Fig. 2f demonstrates the process when the input power gradually increases. As more charge carriers participate in the amplification, less of them get involved in other physical processes. Our observations on the spectrometer also provide circumstantial evidence. As the input power gradually increases, the intrinsic luminescence summit of the VCSEL gradually diminishes until it totally disappears. As a result, less heat is generated from non-radiative recombination, which will lower the local temperature of the active region and widen the band gap of the quantum well material, and finally results in a significant blue shift in the gain spectrum. Similarly, the decrease in temperature could also change the equivalent refractive index of the active region, resulting in a blue shift in the spontaneous emission peak of VCSEL, just not as much as the gain peak, which is already clearly shown in Fig. 2e. Thus, the non-linear amplification could be obtained as long as we set the input wavelength slightly on the blue side of the gain peak ($\lambda_{in} < \lambda_{gain}$). As the input power increases, the blue shift of the gain peak results in less detuning ($\Delta\lambda$), which leads to greater amplification. Subsequently, this positive feedback process reaches electrical saturation, which is limited by the speed of electrical injection. This is the reason why the input-output transfer function exhibits a characteristic sigmoidal shape.

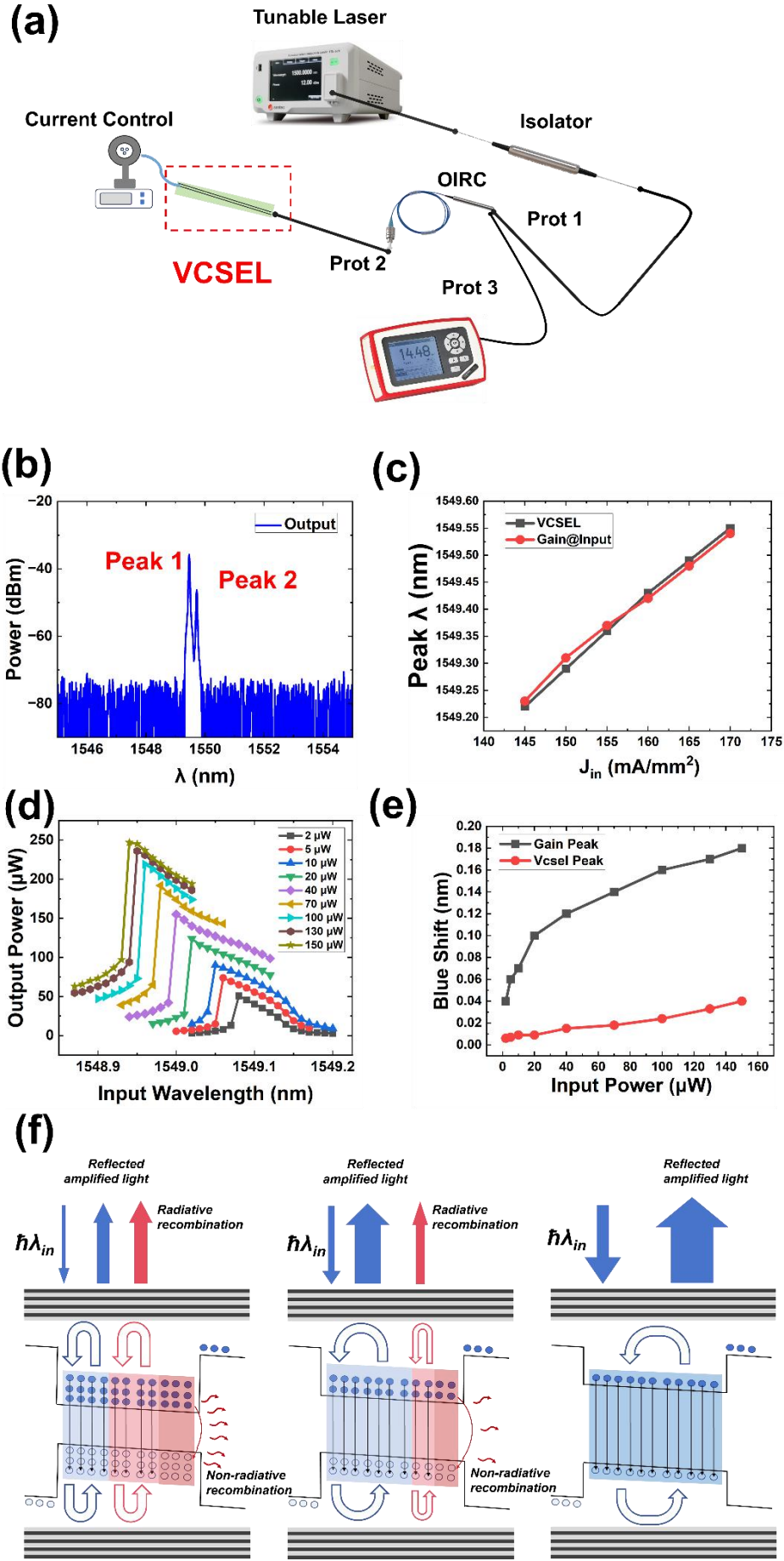

**Fig. 2 Tests and principle analysis of reconfigurable nonlinear computing device based on VCSEL.** (a) Test system: the device receives and transmits information through optical fiber to realize nonlinear operation function. The optical fiber system includes tunable laser, isolator, 3-ports circulator, spectrometer and VCSEL. (b) Output spectral characteristics: Spontaneous radiation peak and injected light's amplification peak are observed. (c) Gain spectrum peak λ and spontaneous radiant luminescence peak λ with different bias currents. (d) Distribution of gain peak spectrum at different input power. (e) Blue shift of the gain peak λ and spontaneous emission peak λ with increasing injected power. (f) The symmetry breaking of VCSEL device leads to the generation of nonlinearity. As injected power increases, the intensity of stimulated amplification will increase, at the same time, the radiation recombination and non radiation recombination gradually weaken to disappear completely.

In the test, different bias current densities are set within the sub-threshold range, as shown in Fig. 3a. Weak injected light are used to determine the initial gain spectra of VCSEL under different bias current densities. We then individually set various levels of wavelength detuning (Δλ) at specific current densities, as depicted in Fig. 3b, 3c, and 3d. Generally, the smaller the Δλ is, the earlier the onset of nonlinearity in the transfer function is observed (referred to as nonlinear threshold power in the following). In addition, the nonlinear threshold power is significantly influenced by the current density. As illustrated in Fig. 3b and 3c, under the condition of a constant wavelength detuning (Δλ=-0.11 nm), a reduction in the injection current density from 165 mA/mm² to 160 mA/mm² results in a modulation of the nonlinear threshold power from 60 μW to 135 μW. Consequently, an impressive tuning sensitivity of 15 μW*mm²/mA is realized. Furthermore, in comparison, it is easy to see that the nonlinear curve at 165 mA/mm2 is steeper. This sharp transition is attributed to the narrow gain peak at this current density, which enhances the positive feedback mechanism. In conclusion, by precisely controlling the injected current density and wavelength detuning (Δλ), we can achieve a specific nonlinear transmission curve, which enables the realization of reconfigurable nonlinear functionalities.

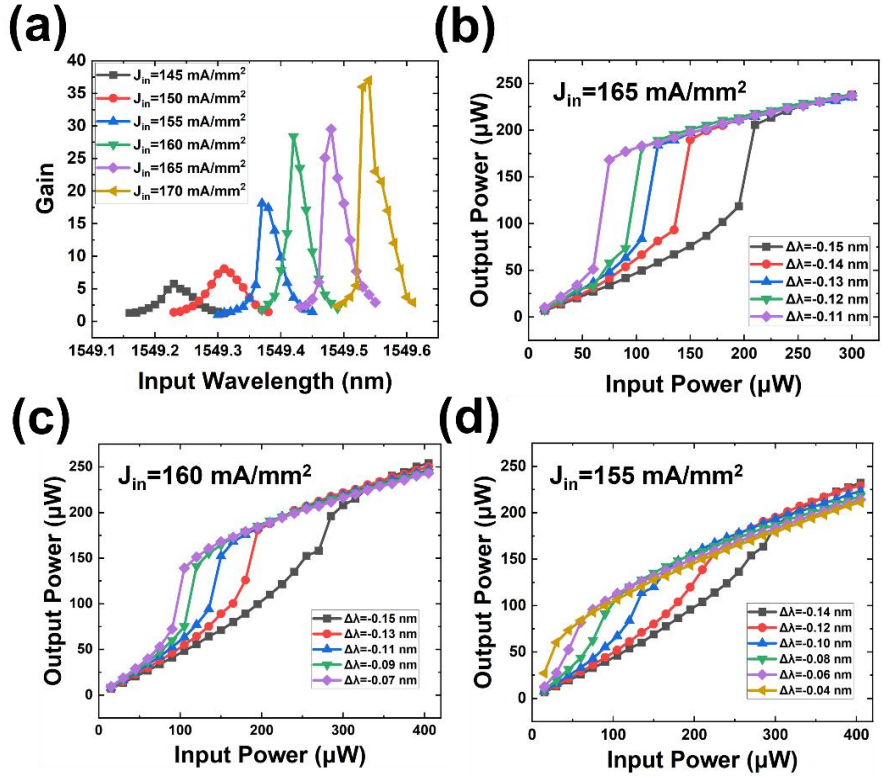

**Fig. 3 Test results of reconfigurable nonlinear transmission characteristics of the device operating at low optical power (ranging from merely 3-250 μW ), a high tuning sensitivity of 15 μW*mm$^2$/mA is achieved.** (a) Gain spectrum distribution at different bias current densities. (b) Transmission characteristics of different wavelength detuning (Current density =165 mA/mm$^2$) (c))Transmission characteristics of different wavelength detuning (Current density =160 mA/mm$^2$) (d))Transmission characteristics of different wavelength detuning (Current density =155 mA/mm$^2$).

In order to verify whether the nonlinearity of the device can really be used as an activation unit in the optical neural networks. Simulation experiments are conducted in specific tasks. Firstly, we mathematically fit the test results with formula (2).

$$y = \frac{A}{B + e^{-Cx}} + Dx \qquad (2)$$

Compared to the classical sigmoid function, we added four parameters A, B, C, D to achieve better fittings, where A/B represents the difference before and after the nonlinear threshold, C represents the magnitude of the input power corresponding to the nonlinear threshold, and D represents the linear term, which indicates the portion of the incident light that reflects without amplification. As shown in Fig. 4, the current density is biased at 160 mA/mm$^2$ with different detunings, the red line is the fitted curve and the black scatters are the tested results.

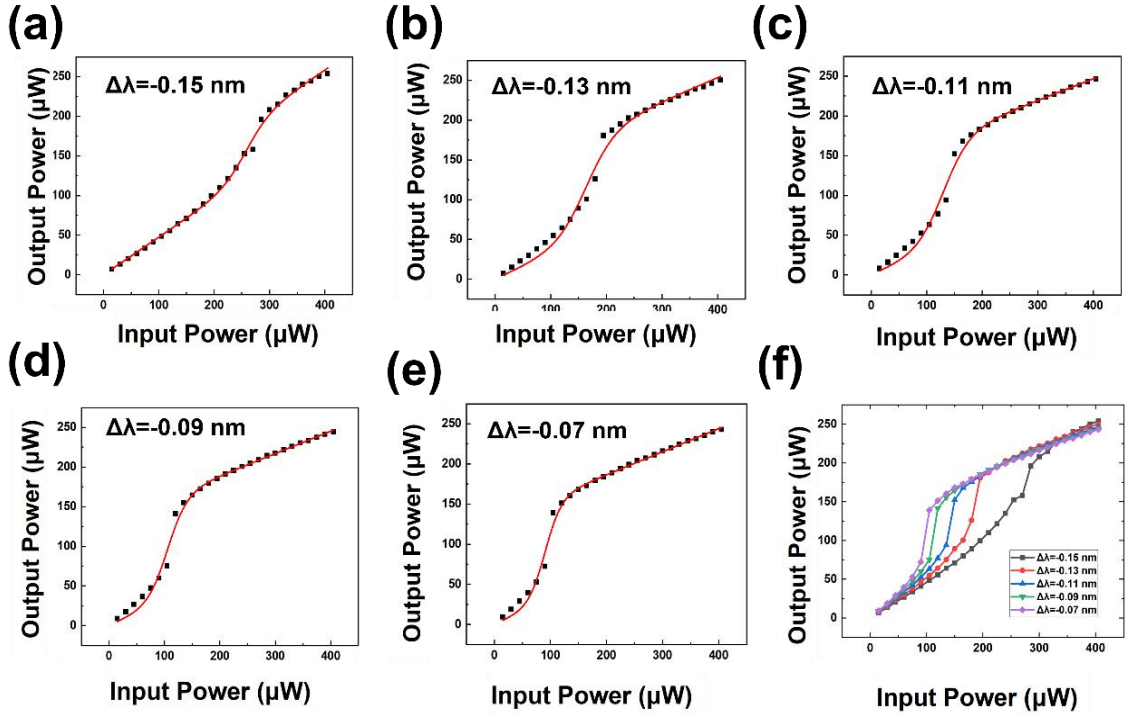

**Fig. 4 Reconfigurable nonlinear mathematical modeling** (a)~(f) Fitting results of nonlinear functions with different wavelength detuning. ($J_{in}$ =160 mA/mm$^2$), the Δλ is set by -0.07 nm, -0.09 nm, -0.11 nm, -0.13 nm and -0.15 nm.

In our simulation, a fully connected neural network tailored for the recognition task of the MNIST dataset was constructed. As illustrated in Fig. 5a, the grayscale values of the 784 pixels serve as input nodes, which are fully connected to the 500 nodes within the first hidden layer. The second hidden layer consists 500 nonlinear activation nodes with full connectivity to 10 nodes of the output layer. For comparison, Leaky ReLU function, fitting function, and linear function are used for activation . Fig. 5b and Fig. 5c respectively show the trends of recognition accuracy and loss with increasing training epochs on the training dataset. Linear neural networks could only achieve an accuracy of 92.1%, and the fitting non-linear functions could improve the accuracy over 97%, which is equivalent to the Leaky ReLU function. To sum up, it is convincing that the nonlinear function provided by our device can function as nonlinear activation units in optical neural networks.

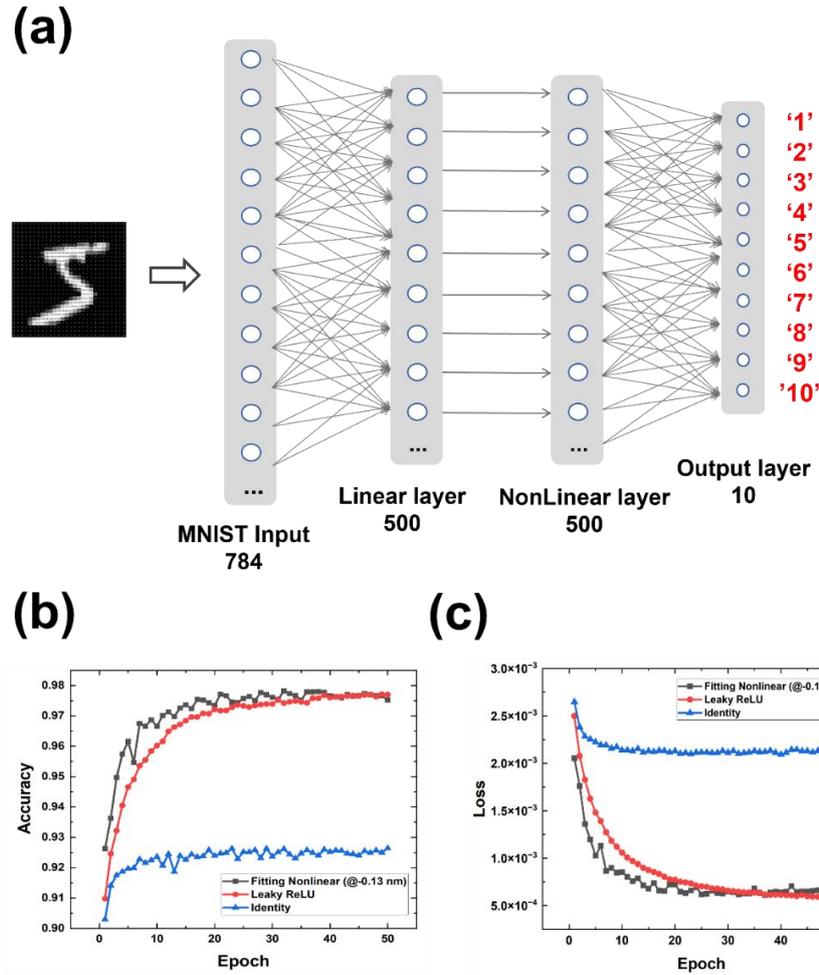

**Fig. 5 The integration of the optical nonlinear function as activation units in the neural network architecture has been shown to significantly enhance the accuracy and expedite the training convergence for the recognition of MNIST dataset of handwritten digits.** (a) Architecture of the neural network, comprising an input layer with 784 nodes, a linear layer with 500 nodes, a nonlinear activation layer with 500 nodes, and an output layer with 10 nodes. (b) Trends in recognition accuracy across training cycles, separately utilizing Leaky ReLU, fitting functions, and linear functions as the nonlinear activation functions. Fitting activation function can achieve recognition accuracy of more than 97%, which is equivalent to Leaky ReLU. (c) Variations in loss over the course of training cycles.

Additionally, the visual targets are not always clear and distinguishable in practice. For humans, the retina has sophisticated mechanisms to preprocess images to enhance the effective features. Our device can be used for artificial retina due to its reconfigurability. As shown in Fig. 6a, we cope with original images by mathematical grayscale processing to obtain under-exposed images, over-exposed images and noisy images. Figure 6c shows three preprocessing nonlinear functions for different types of images, all of which are derived from our fitting results. During the simulation, the initial neural networks is trained on the standard training images, which can achieve 97.2% accuracy, as shown in Fig. 6b. The recognition tasks are then performed on the contaminated images. As shown in Fig. 6d, the accuracy of the recognition has dropped a lot (Noisy: 65.97%, Over-exposed: 19.96% and Under-exposed: 77.8%), because the signal-to-noise ratio of the images has decreased, making their features less significant. After we preprocess the images with the corresponding nonlinear functions, it can be seen that the recognizing accuracy get improved (As shown in Fig. 6d, Noisy: 65.97%-91.86%, Over-exposed: 19.96%-61.72%, Under-exposed: 77.80%-97.00%), because the suitable nonlinear function can drastically improve the SNR of the images and enhance the features of them. In summary, this reconfigurable nonlinearity enables the VCSEL to play variable roles in image preprocessing, making it flexible to cope with images in different practical scenes.

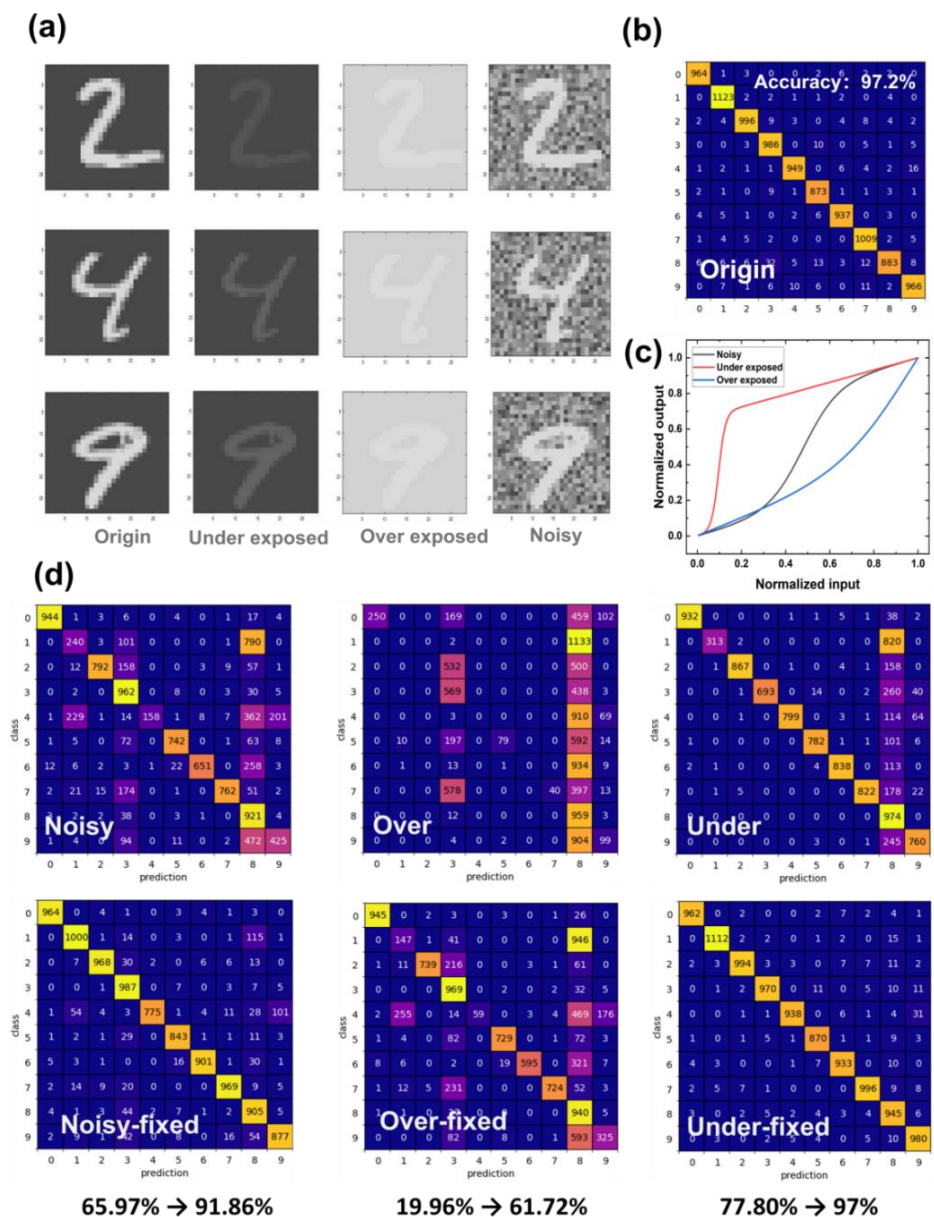

**Fig. 6 The reconfigurability of optical nonlinear functions significantly bolsters the adaptability of neural networks, enabling them to more effectively recognize image information across a variety of scenarios.** (a) We simulated to obtain MNIST

images in different practical scenes, including under-exposed, over-exposed and noisy. (b) Recognition results of two-layer neural network on the original images. (c) Different nonlinear functions for preprocessing images in different scenarios. (d) Recognizing results with trained neural networks applied to images with and without preprocessing. The results indicate that the recognition accuracy is greatly improved after preprocessing with appropriate nonlinear functions. (Noisy: 65.97%-91.86%, Over-fixed: 19.96%-61.72%, Under-fixed: 77.80%-97.00%)

Ultimately, to underscore the superior performance of the device in engineering contexts, a comparative analysis of its performance against other nonlinear activation schemes documented in the literature is conducted. As shown in Table 1, the threshold power, input pattern and reconfigurability are mainly considered. Here the threshold power is defined as the minimum input power at which nonlinear activation can be observed. For the application perspective, this power is expected small enough and adjustable, enabling the device to achieve ideal matching with linear optical computing modules and overall has lower power consumption. In testing, reconfigurable nonlinearity is realized in a large dynamic range (3-250 µW), which could be agilely adjusted by the injected current density (15 µW*mm$^2$/mA) and the wavelength detuning. Furthermore, the nonlinear activations are operable in continuous wave mode, which is compatible with the integration of conventional artificial neural networks. Collectively, these features render our scheme highly valuable for practical applications.

**Table 1 Comparison with reported optical nonlinear activation schemes.**

|  | Device | Wavelength (nm) | Nonlinear threshold | Input model | Reconfigurable |
|---|---|---|---|---|---|
| 30 | DFB-LD | ~ | 26.2 µW | pulse | No |
| 31 | Ge/Si-MRR | 1550 | 0.74 mW | CW | Yes |
| 32 | GST/Si-MRR | 1550 | 0.25~0.63 mW | pulse | Yes |
| 33 | Mxene-MZI | 1180&1490 | 0.33~5.28 mW | pulse | Yes |
| 34 | Ge/Si-OAF | 1549 | 5.1 mW | CW | No |
| 35 | PD-switch | 1550 | 0.2 mW | CW | Yes |
| 22 | PD+MZI | ~ | 0.1 mW | ~ | Yes |
| 36 | Si/SiO$_2$-MZI | 1550 | 2.89 mW | CW | No |
| **This work** | **VCSEL** | **1550** | **3~250 µW** | **CW** | **Yes** |

**DISCUSSION:**
A straightforward optical nonlinear realization scheme employing sub-threshold VCSELs as optical amplifiers is proposed and demonstrated. This scheme enables the injected light to exhibit a nonlinear gain which is shaped like Sigmoid function. By fine-tuning the bias current of the VCSEL and the input wavelength detuning (Δλ), a variety of nonlinear functions can be achieved. Throughout this process, the input power typically varies from a few micro-watts to several tens of micro-watts, which contributes to a notably low energy consumption. In simulation, our fitting functions perform comparably to Leaky-ReLU on the MNIST datasets. Furthermore, its reconfigurability allows for preprocessing images across diverse scenarios. In summary, this device is a concise, energy-efficient and reconfigurable optical nonlinear component, which opens up new prospects of optical neural networks in the future.

## Acknowledgements

All the authors gratefully acknowledge the National Science and Technology Major Project (2021ZD0109903), the National Natural Science Foundation of China (62350002, 62225405, 61991443, 62127814, 62235005, and 61927811), and the Collaborative Innovation Centre of Solid-State Lighting and Energy-Saving Electronics.